\documentclass[twocolumn,showpacs,preprintnumbers,amscd,amsmath,amssymb]{revtex4}
\usepackage{dcolumn}
\usepackage{bm}
\begin{document}
\def\bbox{\vrule height2mm depth0mm width5pt}
\def\theequation{\arabic{equation}}
 \pagenumbering{arabic}
\setcounter{page}{1}
\renewcommand{\d}{{\tt d}}
\newcommand{\F}{\textsf{F}}
\newcommand{\mult}[3]{({\bf{#1}},\,{\bf{#2}},\,{\bf{#3}})}
\newcommand{\disty}{\displaystyle}
\input epsf
\newcommand{\projection}[1]{{\strut #1\,\vrule\,}_{\,\theta=\bar\theta=0}}
\def\DR{\rm I\kern-1.45pt\rm R}
\def\DC{\kern2pt {\hbox{\sqi I}}\kern-4.2pt\rm C}
\newcommand{\bs}{\mbox{\boldmath $\sigma$}}
\newcommand{\bpi}{\mbox{\boldmath $\pi$}}
\newcommand{\bpart}{\mbox{\boldmath $\partial$}}
\newcommand {\cL}{{\cal L}}
\newcommand{\nn}{\nonumber\\}
\newcommand{\p}[1]{(\ref{#1})}
\newcommand{\cZ}{{\cal Z}}
\newcommand{\cH}{{\cal H}}
\newcommand{\bZ}{{\overline Z}}
\newcommand{\bz}{{\bar z}}
\newcommand{\bp}{{\bar p}}
\newcommand{\hp}{{\hat p}}
\newcommand{\bhp}{{\hat{ \overline p}}}
\newcommand{\hQ}{{\hat Q}}
\newcommand{\hH}{{\hat H}}
\newcommand{\bhQ}{{\hat{ \overline Q}}}
\newcommand{\cbZ}{\overline{\cal Z}}
\newcommand {\fc}{{1 + \alpha^2 z {\overline{z}}}}
\newcommand {\fac}{{(1 + \alpha^2 z \overline{z})}}
\newcommand{\cF}{{\cal F}}
\newcommand{\cK}{{\cal K}}
\newcommand{\cV}{{\cal V}}
\newcommand{\cQ}{{\cal Q}}
\newcommand{\cQh}{\widehat{\cal Q}}
\newcommand{\cbF}{\overline{\cal F}}
\newcommand{\bF}{{\overline F}}
\newcommand{\bQ}{{\overline Q}}
\newcommand{\tQ}{{\widetilde Q}}
\newcommand{\tbQ}{\widetilde{\overline Q}}
\newcommand{\tH}{{\widetilde H}}
\newcommand{\bxi}{{\bar\xi}}\newcommand{\bchi}{{\bar\chi}}
\newcommand{\bpsi}{{\bar\psi}}
\newcommand{\I}{{\rm i}}
\newcommand{\ba}{\begin{array}}
\newcommand{\ea}{\end{array}}
\newcommand{\be}{\begin{equation}}
\newcommand{\ee}{\end{equation}}
\newcommand{\bea}{\begin{eqnarray}}
\newcommand{\eea}{\end{eqnarray}}
\newcommand{\bi}{\begin{itemize}}
\newcommand{\ei}{\end{itemize}}
\newcommand{\sfrac}[2]{\mbox{$\frac{#1}{#2}$}}
\newcommand{\D}{{\rm d}}
\newcommand {\bD}{\overline{D}}
\newcommand{\bmth}[1]{\mbox{\boldmath$#1$}}
\title{ Multi-center MICZ-Kepler system,
supersymmetry and integrability}
\author{Sergey
Krivonos$^1$, Armen Nersessian$^{2,3}$ and  Vadim Ohanyan$^{2,4}$
}
\affiliation{
$\;^1$ JINR, Bogolyubov Laboratory of Theoretical Physics, Dubna,
 Russia\\
 $\;^2$  Yerevan Physics Institute  2 Alikhanian Brothers St.,
 Yerevan, 375036,
    Armenia\\
   Yerevan State University, 1 A.Manoogian St., Yerevan, 375025
    Armenia\\
$\;^3$  Artsakh State University,  3 Mkhitar Gosh St.,
  Stepanakert,
Armenia\\
${}^{4}$  Russian-Armenian University, 123 Hovsep Emin St,
 Yerevan,  375051,
 Armenia}
\begin{abstract}

\noindent We propose the general scheme of
incorporation of the Dirac monopoles
into mechanical systems  on the three-dimensional conformal flat
space.
 We found that  any system (without monopoles)
admitting the separation of variables in the elliptic or parabolic
coordinates can be extended to the integrable system with the
Dirac monopoles located at the  foci of the corresponding
coordinate systems. Particular cases of this class of system are
the two-center MICZ-Kepler system in the Euclidean space, the
limiting case when one of the background dyons is located at the
infinity as well as the model of particle in parabolic quantum dot
in the presence of parallel constant uniform electric and magnetic
fields.
\end{abstract}
 \maketitle
\section{Introduction}
 \noindent
The Kepler system is a special one among the finite-dimensional
integrable systems because of its wide applications and the
existence of the hidden symmetry given by the Runge-Lenz vector
\cite{cla}. Many generalizations of the Kepler system were
proposed during the last century, including the Kepler systems on
the spheres and hyperboloids as well as their higher dimensional
generalizations \cite{schr}.
 Zwanziger,  and  McIntosh
and Cisneros independently suggested   generalization of the
Kepler system
 describing the motion of the charged particle in the field of
  Dirac dyon (electrically
charged Dirac monopole), which is presently known as MICZ-Kepler
system \cite{z}. Later on, the similar generalizations have been
found also for the Kepler system on three- and higher-dimensional
 spheres and on the two-sheet hyperboloids  \cite{micsphere}.
Besides the monopole magnetic field, the MICZ-Kepler Hamiltonian
contains the specific centrifugal  potential term  which endows it
with the hidden symmetry of the Kepler system. The essential
differences between the MICZ-Kepler system and the Kepler one are
the lift of the possible range of the orbital angular momentum
\cite{z} and   the modification of the permissible
 dipole transitions \cite{tomilchik}. Indeed, after the incorporation of
the Dirac monopole at the center of spherical coordinates any
spherically symmetric system (without monopole) will undergo only
the minor changes mentioned above, provided the potential term to
be replaced in the following way \cite{mny}
 \be
U_0(r)\to
U_0(r)+\frac{{s}^2}{2Gr^2}. \label{replacement}
 \ee
Here $G(|{\bf r}|)d{\bf r}^2$ is the metrics of the configuration space,
and $s=eg$ is the  ``monopole number"
($e$ is the electric charge of probe particle,
and $g$ is magnetic charge of the monopole).
On the other hand, in the middle of XIX century  Jacobi
established the integrability of the two-center Kepler system and
of  its limiting case when one of the forced centers is placed at
infinity which leads to generation of the homogeneous  potential
field (we shall refer this limiting case as the Kepler-Stark
system because of its obvious relevance to the Stark effect in the
hydrogen atom).
Jacobi also found that these systems admit the
separation of variables in the elliptic and parabolic coordinates,
respectively.
 However, the MICZ-like extensions of these systems,
  as well as of their analogs on the curved spaces,
 were unknown  until now. Respectively, the multi-center
 analog of the MICZ-replacement (\ref{replacement}) for the potentials
 more
 general than the Kepler one was also uknown.

In this paper we define  the ``multi-center MICZ- system" as follows
 \be {\cal H}=\frac{({\bf p}-e{\bf A})^2}{2G({\bf r})}
 + \frac{e^2f^2}{2G({\bf r})}+U_0({\bf r})
\label{multmicz}\ee where ${\bf A}$ is a superposition of the
vector potentials of the Dirac monopoles located at the  points
${\bf a}_I$ (including the vector potential of the Dirac monopole
located at infinity) and $f$ is the potential of the
corresponding   magnetic field: \be
\begin{array}{c}
 {\bf
A}=\sum_{I}{g_I}{\bf A}_{D}({\bf r}-{\bf a}_I) +
\frac{{\bf B}_0 \times {\bf r}}{2},\\[5mm]
 f =\sum_I\frac{g_I}{|{\bf r}-{\bf a}_I|} + {\bf
B}_0{\cdot\;\bf r}.
\end{array}\label{f}\ee
Here  $ {\bf A}_D=\frac{{\bf r}\times {\bf n}}{r( r-{\bf nr})}$, $
{\bf n}^2=1$ is  a vector potential of the Dirac monopole. The
unit vector ${\bf n}_I$ and the constant  $g_I$ determine,
respectively,  the singularity line and the magnetic charge  of
the $I$-th Dirac monopole whereas the constant ${\bf B}_0$
corresponds to  the   magnetic field of the monopole located at
infinity. Its magnitude is given by the expression ${\bf B}_0/G$.
Analyzing the two-monopole configuration (including the monopole
located at infinity)  we arrive at the following conclusion:

{\sl  The  system (without monopoles) admitting separation of
variables in elliptic/parabolic coordinates
results in the separable two-center
 MICZ-system (\ref{multmicz}) with the Dirac monopoles placed at the foci
 of elliptic/parabolic coordinates.}

In this way we obtain the integrable two-center
 MICZ-Kepler system  and the MICZ-Kepler-Stark system.
The  MICZ-Kepler-Stark system is of the special importance as
 it
describes  the one-center MICZ-Kepler system  subjected to
the parallel constant uniform electric and magnetic fields
 with some special confining
potential. One can add the specific oscillator potential
to this system and obtain the integrable model describing
the dyon interacting with the  the parallel constant uniform
electric and magnetic fields in the  parabolic quantum dot.
Switching off the magnetic charge of the dyon
and the constant uniform electric field
 we  arrive at the  integrable model of quantum dot
 considered in the context of condensed matter
 physics few years ago \cite{rashid}.
It is noteworthy  that integrability of the  systems with one- and
two-center Taub-NUT metrics has been established, respectively, in
\cite{gm,horvathy} and \cite{gibbons0}.

The system defined by Eq.(\ref{multmicz}) admits
the  ${\cal N}=4$ supersymmetric extension at the following choice
of one-parametric family of potentials \be
U_{0}=\frac{\kappa}{G}\left(\sum_{I}\frac{g_I}{|{\bf r}-{\bf
a}_I|} + {\bf B}_0{\cdot\;\bf r}\right)+\frac{\kappa^2}{2G}, \quad\kappa=const .
\label{sup}\ee
It was constructed in 2003 by Ivanov and
Lechtenfeld \cite{IL}.
On the Euclidean space it  results in
the multi-center  MICZ-Kepler system with the
background dyons which have the same ratio of the electric and
magnetic charges.

\section{Two-center system: integrability}\noindent
In this Section we show that if the system (without monopoles)
admits the separation of variables in elliptic/paranolic
coordinates then the corresponding  two-center MICZ system
(\ref{multmicz})  with the Dirac monopoles placed at the foci of
elliptic/parabolic coordinate system is also separable in these
coordinates. Particularly, since two-center Kepler system is
separable in elliptic coordinates \cite{cla} the corresponding
two-center MICZ-Kepler system will be also separable in that
coordinate systems (i.e. integrable).

  Let us suppose that the two static monopoles with magnetic
  charges $g_1 $ and $g_2 $ are fixed on $z$-axis at the
  points ${\bf a}=(0,0,a)$ and $-{\bf a}=(0,0,-a)$.
  In the symmetric gauge when Dirac string goes
  along whole $z$-axis the vector potential of the monopole in
  spherical coordinates has the form $A_r=A_\theta =0$,
  $A_{\varphi}=g\cos \theta$.
Thus, in the spherical coordinates the Hamiltonian of
 two-center MICZ-Kepler system  looks as follows
  \be
  \begin{array}{c}
 {\mathcal{H}}=\frac{p^2_r}{2}+\frac{p_\theta^2}{2r^2}+
 \frac{\left( { p}_{\varphi}-s_1\cos \theta_1-s_2 \cos \theta_2 \right)^2}{2r^2 \sin^2 \theta}+\\
 +\frac{1}{2}\left(
\frac{s_1}{r_1}+\frac{s_2}{r_2}\right)^2+ {\cal
U}(r,\theta,\varphi ),\end{array} \label{ham_p-phi(6)}
  \ee
  where ${\mathbf{p}}=(p_r,p_\theta,p_\varphi)$ is the canonical momentum of the system
   and ${\cal U}$ is the appropriate  potential.
Now we turn to
the   elliptic coordinates $(2\xi=(r_1+r_2)/{a}, \; 2\eta=(r_1-r_2)/a)$.
In these coordinates the  Hamiltonian reads
 \begin{eqnarray}
  {\mathcal{H}}=\frac{ \left(\xi^2-1 \right)p_\xi^2+\left(1-\eta^2 \right)p_\eta^2 }{2a^2\left(\xi^2-\eta^2
  \right)}
  +\mathcal{U}_{s_1,s_2}(\xi,\eta,\varphi ), \label{ham_el(15)}
  \end{eqnarray}
  where
  \be\begin{array}{c}
\mathcal{U}_{s_1,s_2}(\xi,\eta ,\varphi )\equiv
\mathcal{U}(\xi,\eta,\varphi )+\\[3mm]
+\frac{1}{2a^2\left(\xi^2-\eta^2
\right)}\left(\frac{{ p}_{\varphi}^2+s_+^2-2p_\varphi s_-
  \xi}{\xi^2-1}
  +\frac{{ p}_{\varphi}^2+s_-^2-2p_\varphi s_+
 \eta}{1-\eta^2}\right),
\end{array} \ee
and $s_{\pm}\equiv s_1 \pm s_2$.
The underlying system (without monopoles) admits  the separation
of variables in elliptic coordinates if its potential term has the
  following form \cite{cla}:
  \be
 {\mathcal{U}}(\xi,\eta)=\frac{{\ V}(\xi)+
 {\ W}(\eta)}{\xi^2-\eta^2}.
 \label{pot_sep(16)}
  \ee
  In this case the system with monopoles  admits the separation of variables too.
 If one put for the generating  function
  $
   S={p}_{\varphi} \varphi +S_1(\xi)+S_2(\eta) -Et$,
    then we get the following Hamilton-Jacobi equations
   \be\begin{array}{c}
   \left(\xi^2-1 \right)\left(\frac{dS_1}{d\xi}
   \right)^2+{\cal V}(\xi)-2a^2 E \left(\xi^2-1 \right)=n,\\
\left(1-\eta^2 \right)\left(\frac{dS_2}{d\eta}
   \right)^2+{\cal W}(\eta)-2a^2 E \left(1-\eta^2 \right)=-n, \end{array}\label{Ham-Jac(19)}
   \ee
   where
   \be
   \begin{array}{c}
   {\cal V}(\xi)=\frac{{p}_{\varphi}^2+s_+^2-2{p}_\varphi s_-
  \xi }{\xi^2-1}+2a^2V(\xi),\\[3mm]
  {\cal W}(\eta)=\frac{{p}_{\varphi }^2+s_-^2-2{p}_\varphi s_+
 \eta }{1-\eta^2}+2a^2W(\eta),
\end{array}\label{vw}\ee
and $n$ is separation constant.
 The  two-center MICZ-Kepler  system with arbitrary values of electric and magnetic charges
 belongs to this class: its potential  is given by the expression ${\cal
U}=eq_1/r_1+eq_2/r_2 $, and
 could be represented in the form (\ref{pot_sep(16)}).

From (\ref{Ham-Jac(19)})-(\ref{vw}) we can immediately get the
explicit expression for the quadratic (on momenta) constant of
motion, which is responsible for the separation of variables \cite{cla}.
That is
 \be
{\mathcal{I}}=\frac{\eta^2[(\xi^2-1)p^2_\xi+{\cal
 V}(\xi)]+
\xi^2\left[(1-\eta^2)p_\eta^2
 -{\cal W}(\eta)\right]
}{\xi^2-\eta^2}.
\ee
 Though we restrict ourselves to the system in the Euclidean space
 it is clear from the consideration
 that our conclusion remains valid for the systems
 with any metric which admits the separation
 of variables in elliptic coordinates.

   \section{ MICZ-Kepler-Stark(-Zeeman) system}
   \noindent
   The systems considered above  have important limiting case.
   If we move one of the background monopoles to infinity then
    it will generate the  constant uniform
    magnetic field.
 Similar to previous case we are going to  show that if the
system (without monopoles) allows the separation of variables in
the parabolic coordinates
 then  it  admits
the separation  of variables also in the presence of the monopoles
placed at the foci of parabolic coordinates.

In the spherical coordinates the vector-potential of the constant
uniform magnetic field,
   assumed to be in $z$-direction, is
$A_r=A_\theta=0$, $A_\varphi=\frac{B}{2}r^2\sin^2\theta
  $.
 The Hamiltonian reads:
   \be\begin{array}{c}
 {\mathcal{H}}=\frac{p^2_r}{2}+\frac{p_\theta^2}{2r^2}+
  \frac{\left(p_\varphi-s\cos\theta-\frac{1}{2}eBr^2\sin^2\theta
   \right)^2}{2r^2\sin^2\theta}+\\[3mm]
   +\frac{1}{2}\left(\frac{s}{r}+eBz \right)^2 +{\cal U}(r,\theta,\varphi ).
   \end{array}\label{ham_par(22)}
   \ee
Let us choose the  parabolic coordinates given by the relations
   $\xi=r+z,\; \eta=r-z$,
   and suppose that the potential
    ${\cal U}$ is  separable in parabolic coordinates \cite{cla}
\begin{eqnarray}
{\mathcal{U\left(\xi, \eta
\right)}}=\frac{U(\xi)+V(\eta)}{\xi+\eta}. \label{par_pot_sep(28)}
\end{eqnarray}
Then the Hamiltonian could be presented as follows \be
{\mathcal{H}}=\frac{ 4\xi p_\xi^2+4\eta
p_\eta^2
+{\mathcal{V}}(\xi)+{\mathcal{W}}(\eta)}{2(\xi+\eta)}
-\frac{p_\varphi e B}{2}
,
\label{par_ham(27)} \ee
where \be\begin{array}{c}
{\mathcal{V}}(\xi)=\frac{(p_\varphi+s)^2}{\xi}
 +3seB\xi +\frac{e^2B^2\xi^3}{4} +2 V(\xi),\\[3mm]
 {\mathcal{W}}(\eta)=\frac{(p_\varphi-s)^2}{\eta}
 -3seB\eta +\frac{e^2B^2\eta^3}{4}+2 W(\eta). \end{array}\ee
The separability of the system is obvious. The constant of motion
which is responsible for the separability of variables
 looks as follows
\be {\mathcal{I}}=\frac{4\xi\eta (p^2_\xi-p^2_\eta) +\eta{\cal
V}(\xi) -\xi{\cal W}(\eta)}{\xi+\eta}. \ee Notice that similar to
elliptic case, our conclusions concerning integrability remains
valid
 also for the systems with non-constant metric
  $G$ which admit the separation of variables
 in parabolic coordinates.

The important particular case of the systems under consideration
is the Jacobi problem when the potential is the sum of the Coulomb
potential and  the potential of the constant uniform electric field
parallel to the magnetic one (Kepler-Stark system):
 ${\cal U}=eq/r-e{\bf E}$.
 In that case the monopole placed at infinity, generates the
 constant uniform magnetic field ${\bf B}_0$ which is parallel to
 the electric one: ${\bf B}_0 \|{\bf E}$.
 In that case the system could be viewed
 as  the MICZ-Kepler-Stark system.
  Its  Hamiltonian looks as follows:
\be\begin{array}{c} {\cal H}=\frac{({\bf p}-s{\bf A}_D- e{\bf
B_0\times
r}/2)^2}{2}+\frac{s^2}{2r^2}+\frac{eq}{r} -e{\bf Er}+\\[3mm]
 +\frac{es{\bf B}_0{\bf
r}}{r}+\frac{(e{\bf B_0r})^2}{2}.\end{array} \label{ex}\ee Hence,
the two-center MICZ-Kepler-Stark system can be interpreted as the
(one-centered) MICZ-Kepler system in the parallel constant uniform
   electric and magnetic fields  with  some additional potential.
    Due to this additional correction
   the system becomes integrable, in contrast to the
    MICZ-Kepler system in
   the constant uniform electric and magnetic fields.

One may also extend the Hamiltonian (\ref{ex}) with  the additional
oscillator potential of the type (\ref{par_pot_sep(28)})
\be V_{{\rm
conf}}=
\frac{\omega^2_0(r^2+3z^2)}{2}=\frac{\omega^2_0(\xi^3+\eta^3)}{2(\xi+\eta)}.
\ee In this case putting $s=0$ we obtain the Coulomb particle in the
parallel electric and magnetic fields  placed in the axially
symmetric quantum dot with confining potential of parabolic type
with the frequencies
 $\omega_x=\omega_y\equiv \omega_0$, $\omega_z=2\sqrt{\omega_0^2+(eB_0/2)^2}$ .
The similar integrable system without electric  field
 has been recently proposed in the
context of condensed matter physics as a model of two-electron
quantum dot \cite{rashid}.

\section{Supersymmetry}\noindent
As it was mentioned in Introduction, the MICZ-system
(\ref{multmicz}) with the potential (\ref{sup}) admits the ${\cal
N}=4$ superextension, the corresponding system was constructed in
\cite{IL} (for earlier work devoted to
 supersymmetrization of
 (one-center) MICZ-Kepler system and related issues see
 \cite{susymic} and refs therein).
In the Hamiltonian approach
it
is described
 by the canonical Poisson brackets $\{x^n,p_m\}=\delta^{n}_m$,
$\{\chi^\alpha,\bchi_\beta\}=\imath\delta^\alpha_\beta $
 and by the following  Hamiltonian
\be
\begin{array}{c}{\cal H}_{SUSY}=\frac{\bpi^2}{2G}+
\frac{e^2(f +\kappa)^2}{2G}+  {e{\bf
B\Lambda}}\\[3mm]
+\frac{(\bpi\times{\bf \Gamma})\cdot {\bf
\Lambda}+e(f+\kappa){\bf\Gamma \Lambda}}{G} +
\left(\frac{\bpart\cdot {\bf \Gamma}+2{{\bf\Gamma}^2}}{2G}
\right){(\chi\bchi )^2}.
\end{array}
\label{hamiltonian}\ee
The supercharges look as follows
 \be Q_\alpha=\frac{1}{{\sqrt
G}}\left[\left(\imath\widehat{\bpi}+
2(\chi\bchi)\widehat{{\bf\Gamma}})\bar\chi\right)_\alpha
-e(f +\kappa) \bar\chi_\alpha \right].
\label{hsupercharge}
\ee
 Here we use the notations $\bpi\equiv{\bf p}-e{\bf A}$, and
 \be {\bf
B}\equiv\frac{{\bpart\times {\bf A}}}{G},\;
  {\bf \Gamma}\equiv\frac{{\bpart }\log G}{2},\;
{\bf\Lambda}\equiv{\chi\bs\bchi}\;
.
 \label{mf}\ee
 One can see that  ${\bf B}$ is the magnitude of the
external magnetic field which emerged in the system and
${\bf\Lambda }$ defines  the  spin matrices. The third term
in the Hamiltonian may be interpreted  as Zeeman energy.
To quantize the system one has to replace the odd variables
$\chi^a$ by four-dimensional Euclidean
 gamma-matrices $ \hbar({\widehat\gamma}^a+\imath{\widehat\gamma}^{a+2})/\sqrt{2}$
 and $\bchi_a$ by their Hermitean conjugates,
 and  $(\chi\bchi)^2$ by
 ${\widehat\gamma}_5$ matrix. The momentum will be quantized as follows
 $\bpi\to -\imath\hbar\bpart-e{\bf A} $.

 In the Euclidean space the bosonic part of potential looks as follows
 \be
\begin{array}{c}
\frac{e^2(f +\kappa)^2}{2}=\sum_{I,J} \frac{s_Is_J}{2r_Ir_J} + \sum_I
\frac{ee_I}{r_I}+ e{\bf E}_0{\bf r} + {\cal
E}_0+\\[3mm]
 +{e\bf B}_0{\bf
\cdot r}\left(\sum_I\frac{s_I}{r_I}\right)+
\frac{(e{\bf B}_0{\bf\cdot\; r})^2}{2} ,
\quad r_I\equiv |{\bf r}-{\bf a}_I|,\end{array}
\label{s1}\ee where $ s_I\equiv  eg_I$, $e_I\equiv \kappa s_I$, ${\bf
 E}_0\equiv e\kappa{\bf B}_0$, ${\cal E}_0\equiv{e^2\kappa^2}/{2}$.
One can obviously interpret   $e_I$ as the electric charge of the $I$-th
monopole, ${\bf E}_0$ and ${\bf B}_0$ as the magnitudes
of the constant uniform electric and magnetic fields which are
 parallel to each other,  and ${\cal E}_0$ as a ground
state energy of the system. Thus, the constant  $\kappa$ provides the
background monopoles  with electric charges,
i.e. turns them into dyons. All these dyons
 have the same ratio of the electric and magnetic charges
${e_I}/{g_I}= |{\bf E}_0|/|{\bf B}_0| = e\kappa$ and therefore
obey trivial  Dirac-Schwinger-Zwanziger charge quantization
condition:
  \begin{eqnarray}
 s_{IJ}\equiv e_I g_J-e_J g_I=0,\quad s_{I\infty}=0.
  \label{DSZ}
  \end{eqnarray}
One should remind  that   in the general case one have
$s_{IJ}/(2\pi\hbar) \in {\mathbb{Z}}$. Notice, that this system
can have not only repulsive  Coulomb potential but an attractive
one as well: the presence of the ``ground state energy" term
${\cal E}_0$ makes the Hamiltonian positive in the whole range.

{\sl Hence,  on the Euclidean space ($G=1$)
the Hamiltonian  (\ref{hamiltonian}) defines
 ${\cal N}=4$ supersymmetric multi-center MICZ-Kepler system
with the background dyons satisfying the condition (\ref{DSZ}).
 In the case of single background
dyon (MICZ-Kepler system) there is
  no   restriction on
the permissible values of electric
and magnetic charges.}

Unfortunately, this parametric family of Hamiltonians in
 case of sphere and hyperboloid does not yield the corresponding
MICZ-Kepler counterpart. Let us remind that the conformal flat
metrics on the sphere and hyperboloid  is defined by the factor
$G(r)=4(1\pm {\bf r}^2)^{-2}$ and the Coulomb
 potential is given by the expression
 $U_0=\gamma (1\mp {\bf r}^2)/r$
 (upper sign corresponds to the sphere, and lower-to the hyperboloid;
 we assume that sphere/hyperboloid has unit radius).

However, there is another important case,  $G=(f+\kappa)$, which
 describes the motion
of the probe particle in the background of
well-separated BPS monopoles/dyons \cite{gm}.
In this case the potential looks as follows
$U= e^2{\bf Br}+\sum_i e^2g_I/r_I+\kappa$. Hence,  similar to the
Euclidean case, the electric charges are proportional to the
magnetic ones. Moreover, this system admits further extension from
${\cal N}=4$ to ${\cal N}=8$ supersymmetry.
Indeed, the constant
$e$ in (\ref{hamiltonian}) can be considered
  as a momentum $p_\phi\equiv e$ conjugated to some
cyclic variable $\phi$ \cite{KS}. Then the system may be viewed as
 the free particle moving
in the four-dimensional space endowed with the metric \be
ds^2=G({\bf r})d{\bf r}^2+ \frac{G({\bf r})(d\phi+{\bf A}d{\bf
r})^2}{(f +\kappa)^2}, \label{hk}\ee where  $\bpart\times {\bf
A}=-\bpart (f+\kappa)$. At $G=f+\kappa$
 the metric  (\ref{hk}) possesses the
  hyper-K\"ahler structure.  For $f$ given by the expression (\ref{f})
it yields  the multi-center Taub-NUT metric.
Thus, the initial
  ${\cal N}=4$ supersymmetry can be extended
to the ${\cal N}=8$  one by adding a proper number of fermionic
degrees of freedom \cite{korea}.

The Euclidean and Taub-NUT systems are related to each other.
Indeed, the energy surface of the Taub-NUT system ${\cal
H}_{Taub}=E_{Taub}$ can be written as follows $
{\bpi^2}/{2}+{e^2(f+\kappa)^2}/{2}-(f+\kappa)E_{Taub}=0$. Taking
into account the expression (\ref{f}) we arrive at the energy
level of the multi-center MICZ-Kepler system with the background
dyons obeying the condition (\ref{DSZ}). It is  clear  that there
is no such a simple correspondence between supersymmetric versions
of these systems.

\section{Conclusion}
Let us emphasize the main statements of this paper:
\begin{itemize}
\item We presented the multi-center  MICZ-Kepler system which
describes the motion of the
probe electrically charged particle in multi-dyon background.

\item We proved the exact solvability of the  two-center MICZ-Kepler  and
 of the MICZ-Kepler-Stark systems.
 \item We extended the MICZ-Kepler-Stark  system  to an
integrable model of the particle interacting with the parallel constant uniform electric and magnetic fields
in the parabolic  quantum dot.
\item We have shown that
the ${\cal N}=4$ supersymmetric mechanics, constructed by Ivanov
and Lechtenfeld, \cite{IL} describes (in the Euclidean space) the
supersymmetric extension of multi-center MICZ-Kepler system with
the same ratio of the electric and magnetic charges of background
dyons.
\end{itemize}
The results listed above are more general because
they embrace a wide range of system and potentials including
important cases of curved configuration spaces. The direct
relation of the MICZ-Kepler-Stark system with the problems
arising in the physics of nanostructures can receive further
development and applications especially in the quantum-mechanical
context. The quantum-mechanical description of MICZ-Kepler and
MICZ-Kepler-Stark systems, as is known, has much in common with
the Hamilton-Jacobi formalism presented in this paper. We intend
to study this issue in our further publications.

Finally, generalizing the construction considered in the paper we
 suggest an anzats which could describe the few-body
MICZ-Kepler system in a proper way:
 \be\begin{array}{c} {\cal H}=\sum_{I}\left[\frac{1}{2}\left({\bf
p}_I-\sum_Js_{IJ}A_{D}({\bf r}_{IJ})\right)^2+\right.\\[3mm]
\left. +\frac{1}{2}\sum_{J,K}\frac{s_{IJ}s_{IK}}{r_{IJ}r_{IK}}+\sum_J\frac{\alpha_{IJ}}{r_{IJ}}\right],\quad
{\bf r}_{IJ}\equiv{\bf r}_I -{\bf r}_J.\end{array} \ee
Obviously, this system is not
exactly solvable. But, apparently, there should exist the static
configurations of the particles analogous to those well-known
from the celestial mechanics of three-body systems.\\

We are grateful to  Andrey Scherbakov
for his contribution  at the earlier stage of this work. We thank
Calogero Natoli  and Rashid Nazmitdinov for the interest in work
and useful comments. A. N. acknowledge the International Center
for Theoretical Physics (Trieste), where this work was completed,
for hospitality. The work was supported in part   by the grants
 RFBR-06-02-16684 (S.K.), DFG~436 Rus~113/669/03 (S.K.), NFSAT-CRDF  ARPI-3228-YE-04 (A.N.),
 and  INTAS-05-7928.

\end{document}